# List-Mode PET Image Reconstruction Using Deep Image Prior

Kibo Ote, Fumio Hashimoto, Yuya Onishi, Takashi Isobe, and Yasuomi Ouchi

*Abstract*—List-mode positron emission tomography (PET) image reconstruction is an important tool for PET scanners with many lines-of-response and additional information such as time-of-flight and depth-of-interaction. Deep learning is one possible solution to enhance the quality of PET image reconstruction. However, the application of deep learning techniques to list-mode PET image reconstruction has not been progressed because list data is a sequence of bit codes and unsuitable for processing by convolutional neural networks (CNN). In this study, we propose a novel list-mode PET image reconstruction method using an unsupervised CNN called deep image prior (DIP) which is the first trial to integrate list-mode PET image reconstruction and CNN. The proposed list-mode DIP reconstruction (LM-DIPRecon) method alternatively iterates the regularized list-mode dynamic row action maximum likelihood algorithm (LM-DRAMA) and magnetic resonance imaging conditioned DIP (MR-DIP) using an alternating direction method of multipliers. We evaluated LM-DIPRecon using both simulation and clinical data, and it achieved sharper images and better tradeoff curves between contrast and noise than the LM-DRAMA, MR-DIP and sinogram-based DIPRecon methods. These results indicated that the LM-DIPRecon is useful for quantitative PET imaging with limited events while keeping accurate raw data information. In addition, as list data has finer temporal information than dynamic sinograms, list-mode deep image prior reconstruction is expected to be useful for 4D PET imaging and motion correction.

*Index Terms*—Deep neural network, image recon-struction, list-mode, positron emission tomography, unsupervised learning.

## I. INTRODUCTION

POSITRON emission tomography (PET) is an in-vivo imaging tool with a wide range of clinical applications such as oncology, cardiology, and neurology [1]. In PET scan, firstly a radiotracer is injected into a subject. Then, the radiotracer emits a positron, and an annihilation between the positron and electron generates two 511 keV gamma rays. The PET detector ring catches two gamma rays as a coincidence event (i.e., the two annihilation photons (or gamma rays) are detected within a time coincidence window). The coincidence event indicates that an activity is on the line-of-response (LOR), but it cannot be determined where the activity is on the LOR. Hence, an image reconstruction process is required for tomographic imaging.

Positron emission tomography image reconstruction is a challenging task because the number of events is limited, and it is also an ill-posed inverse problem. Regularization using prior information about PET images is often used to suppress the ill-posedness and statistical noise. For example, the spatial smoothness of an image is used to regularize the PET image reconstruction through Gibbs priors [2], [3]. In addition, anatomical information from X-ray computational tomography (CT) or magnetic resonance imaging (MRI) can be used for edge-preserving regularization of PET image reconstruction [4]-[6]. Besides these conventional algorithms, some deep learning-based approaches have recently been proposed for PET image reconstruction and provide excellent performance [7]-[10].

One strategy of deep learning-based PET image reconstruction is to learn the image reconstruction from sinograms through large amounts of training datasets [11], [12]. For example, [13] proposed DeepPET, which performs a direct mapping from a sinogram space into reconstructed image space, using a fully convolutional neural network (CNN). It is easy to handle with CNN because both the sinogram and reconstructed image are represented by the tensor. In another strategy, [14] proposed a regularization method that uses a CNN representation as prior information. However, these deep learning-based methods still require large amounts of datasets for network training. In recent years, unsupervised learning has attracted attention for PET image reconstruction, particularly deep image prior (DIP) [15], which uses the CNN architecture itself as a regularization, is used as a new regularization tool for PET image reconstruction because it requires no prior training data [16]-[18]. However, the number of bins of a sinogram is dependent on the number of LORs, and further increases in orders of magnitude if the PET scanner has the ability to measure additional information, such as time-of-flight (TOF) and depth-of-interaction (DOI). This is because the TOF and DOI increase the dimension of the sinogram and increase the number of LORs. Hence, the sinogram-mode PET image reconstruction becomes memory intensive and sometimes infeasible, when the PET scanner has many LORs and additional information.

List-mode PET image reconstruction is the promising solution for TOF- and DOI-PET image reconstruction when the number of events is less than the number of LORs or when the sinogram is too large to fit in memory. In addition, the list-mode method is efficient for dynamic PET image reconstruction because the number of events per frame becomes low [19] and is suitable for motion correction because the list data holds fine temporal information. Hence, improving the list-mode PET image reconstruction is important for advanced PET scanners. However, the incorporation of deep learning techniques to list-mode PET image reconstruction has not advanced because list data is a sequence of bit codes and is not suitable for processing by CNNs.

To surpass the above barriers, in this study, we propose list-mode PET image reconstruction using DIP which is the first trial to integrate list-mode PET mage reconstruction and CNN. The proposed method, called LM-DIPRecon, alternatively iterates list-mode PET image reconstruction and image processing by DIP, according to an alternating direction method of multipliers (ADMM) [20]. The proposed LM-DIPRecon is a strong algorithm for PET image reconstruction while keeping accurate raw data information although a sinogram-based reconstruction loses some information to calculate practical and manageable sinograms. We evaluated the proposed LM-DIPRecon method using Monte-Carlo simulation and clinical data. The LM-DIPRecon method reduces noise and maintains the contrast compared with post-processing by the DIP , and sinogram-based DIPRecon methods.

## II. RELATED WORKS

In this section, we introduce list-mode PET image reconstruction and DIP.

*A. List-mode PET image reconstruction*

The list data is formulated as,

$$U = \{i(t)|t = 1.2.\cdots,T\}, \quad (1)$$

where $t$ is an index of a coincidence event, $T$ is the number of events, and $i(t)$ is an index of LOR measuring the $t$-th event.

The list-mode log-likelihood function [21] is defined as,

$$L(U|x) = \sum_t \log \sum_j a_{i(t)j} x_j - \sum_i \sum_j a_{ij} x_j, \quad (2)$$

where $x$ is an unknown image, $j$ is an index of a voxel, and $a_{ij}$ represents the contribution of the voxel $j$ to the LOR $i$.

We can reconstruct the image from the list data by maximizing the list-mode log-likelihood function with an iterative algorithm. For fast global optimization, the list-mode dynamic row action maximum likelihood algorithm (LM-DRAMA) has been developed [22], [23]. The reconstructed image $x_{\text{EM}}$ is calculated using LM-DRAMA updates, as follows:

$$x_{j,\text{EM}}^{(k,q+1)} = x_j^{(k,q)} + x_j^{(k,q)} \lambda^{(k,q)} \left( \frac{N_{\text{sub}}}{S_j} \sum_{t \in \text{Sub}_q} \frac{a_{i(t)j}}{y_{i(t)}^{(k,q)}} - 1 \right), \quad (3)$$

$$S_j = \sum_i a_{ij}, \quad (4)$$

$$y_i^{(k,q)} = \sum_j a_{ij} x_j^{(k,q)}, \quad (5)$$

$$\lambda^{(k,q)} = \frac{\beta}{\beta + q + \gamma k N_{\text{sub}}}, \quad (6)$$

where $k$ is the number of main iteration, $q$ is the number of sub-iteration, $N_{\text{sub}}$ is the number of subsets of list data, $S$ is the sensitivity image [24], $y$ is the forward projection of the image, $\lambda$ is a subset-dependent relaxation coefficient, $\text{Sub}_q$ is a subset of list data accessed at the $q$-th sub-iteration, and $\beta$ and $\gamma$ are the parameters of the relaxation coefficient.

*B. Supervised deep learning*

The supervised deep learning approach trains a CNN to map a degraded image to a clean image, and requires large amounts of pairs of degraded and clean images as the dataset.

Supervised deep learning is formulated as,

$$\underset{\theta}{\text{argmin}} \sum_{s \in D} \|f_\theta(x_0^s) - x_{\text{ref}}^s\|^2, \quad (7)$$

Where $f$ is a CNN, $\theta$ is a parameter of the CNN such as the connection weight, and $x_0^s$ and $x_{\text{ref}}^s$ are $s$-th pair of measured degraded image and clean image in the dataset $D$.

*C. DIP*

In contrast to supervised deep learning, an unsupervised deep learning technique requires only the degraded image as the dataset without clean images [25]. Further, the deep image prior requires no prior training data other than the measured degraded image itself [15]. The DIP method optimizes a CNN to map a prior distribution image to a measured degraded image. When there is no prior information, DIP inputs a random image noise to the CNN as the prior distribution image. CNNs have the property of restoring signals faster than noise [15]. By stopping the optimization early, we can obtain a restored image as an output of the CNN.

The DIP method is formulated as,

$$\underset{\theta}{\text{argmin}} \|f_\theta(z) - x_0\|^2, \quad (8)$$

where $z$ is a prior distribution image.

In addition to the network architecture, the prior distribution image is the key factor that decides the performance of the DIP. For example, if we use the CT or MRI as the prior information [26], the recovery of PET images by DIP becomes faster and more accurate than the case of no prior information. For dynamic PET image denoising, we can also use the static PET image as the prior information [27].

## III. METHODOLOGY

In this section, we integrate list-mode PET image reconstruction and DIP into one algorithm using ADMM.

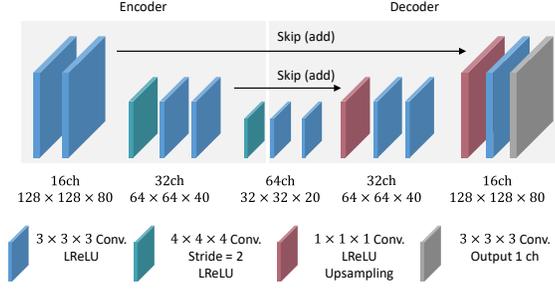

Fig. 1. Network architecture of 3D U-net in this study.

### A. Integration of LM-DRAMA and DIP

To regularize list-mode PET image reconstruction using DIP, we consider the following constrained optimization problem.

$$\max_{x,\theta} L(U|x) \quad \text{s.t.} \quad x = f_\theta(z), \tag{9}$$

where the unknown image $x$ is constrained to the output of the CNN.

The constrained optimization problem in (9) can be rewritten by the augmented Lagrangian method.

$$\min_{x,\theta} \max_g -L(U|x) + g^T(x - f_\theta(z)) + \frac{\rho}{2}\|x - f_\theta(z)\|^2 \tag{10}$$

where $\rho$ is a positive constant and $g$ is a dual variable or Lagrange multipliers. The optimization problem of (10) can be written in a slightly different form [20].

$$\min_{x,\theta} \max_\mu -L(U|x) + \frac{\rho}{2}\|x - f_\theta(z) + \mu\|^2 - \frac{\rho}{2}\|\mu\|^2 \tag{11}$$

where $\mu = g/\rho$ is a scaled dual variable.

The ADMM splits the optimization problem of (11) into the two subproblems and the update of scaled dual variable [20].

$$x^{(n+1)} = \operatorname*{argmax}_x L(U|x) - \frac{\rho}{2}\left\|x - f_{\theta^{(n)}}(z) + \mu^{(n)}\right\|^2, \tag{12}$$

$$\theta^{(n+1)} = \operatorname*{argmin}_\theta \left\|f_\theta(z) - x^{(n+1)} - \mu^{(n)}\right\|^2, \tag{13}$$

$$\mu^{(n+1)} = \mu^{(n)} + x^{(n+1)} - f_{\theta^{(n+1)}}(z), \tag{14}$$

where $n$ is the number of iterations of three steps of the ADMM.

### B. Solving subproblem (12)

The subproblem (12) can be solved by the optimization transfer method [16], [28], [29]. Firstly, to optimize each voxel independently, we construct the surrogate function of $L(U|x)$ as,

$$Q_L(x|x^{(n)}) = \sum_j S_j\left(x_{j,\text{EM}}^{(n+1)} \log x_j - x_j\right). \tag{15}$$

where $x_{\text{EM}}^{(n+1)}$ is an updated image of $x^{(n)}$ by (3).

The $Q_L(x|x^{(n)})$ will suffice the following two conditions from the convergence property of the DRAMA [30].

$$Q_L(x|x^{(n)}) - Q_L(x^{(n)}|x^{(n)}) \le L(U|x) - L(U|x^{(n)}), \tag{16}$$
$$\nabla Q_L(x^{(n)}|x^{(n)}) = \nabla L(U|x^{(n)}). \tag{17}$$

By replacing the $L(U|x)$ with the surrogate function $Q_L(x|x^{(n)})$, we obtain the surrogate objective function of subproblem (12) as,

$$P(x_j|x^{(n)}) = S_j\left(x_{j,\text{EM}}^{(n+1)} \log x_j - x_j\right) - \frac{\rho}{2}\left(x_j - x_{j,\text{base}}^{(n)}\right)^2, \tag{18}$$

$$x_{j,\text{base}}^{(n)} = f_{\theta^{(n)}}(z)_j - \mu_j^{(n)}, \tag{19}$$

where $x_{\text{base}}$ is a base image of the regularization.

The surrogate objective function (18) can be optimized voxel by voxel. Hence, by setting the derivative of equation (18) to be zero, we can obtain the following update equation.

$$x_j^{(n+1)} = \frac{x_{j,\text{base}}^{(n)} - \frac{S_j}{\rho} + \sqrt{\left(x_{j,\text{base}}^{(n)} - \frac{S_j}{\rho}\right)^2 + 4x_{j,\text{EM}}^{(n+1)}\frac{S_j}{\rho}}}{2}. \tag{18}$$

As a consequence, the algorithm of subproblem (12) becomes a combination of LM-DRAMA and regularization by equation (20).

In this study, we set $N_{\text{sub}} = 40$, $\beta = 30$, and $\gamma = 0.1$ for the LM-DRAMA. These are the default settings of LM-DRAMA. In addition, we set $\rho = 0.5$ for the regularization based on the experimental results (Fig. 5).

### C. Solving subproblem (13)

Subproblem (13) can be solved by the DIP with $x_{\text{label}} = x^{(n+1)} + \mu^{(n)}$. In this study, we employ the MR-DIP which uses the MR image as the prior distribution image. In addition, we use the 3D U-net architecture [31] for the MR-DIP because it is suitable for medical image processing, and it may have a good inductive bias for the PET image. Fig 1 shows the network architecture of the 3D U-net in this study. The 3D U-net consists of encoder, decoder, and skip connections.

The encoder extracts the feature maps to generate the PET image from the MR image. The combination of 3D convolution with a kernel size of three and a leaky rectified linear unit (LReLU) is repeated twice before down-sampling. Down-sampling is performed by the combination of 3D convolution with kernel size of four and stride of two and LReLU. At each down-sampling, the number of voxels along $x$, $y$, and $z$ directions are halved and the number of channels of feature map is doubled.

The decoder reconstructs the PET image from the extracted feature maps. The combination of 3D convolution with kernel size of three and LReLU is repeated twice before up-sampling. Up-sampling is performed by the 3D convolution with kernel size of one and LReLU and trilinear interpolation. At each up-sampling, the number of voxels along $x$, $y$, and $z$ directions are doubled, and the number of channels of feature map is halved. The output image is reconstructed by 3D convolution with kernel size of three with a single channel output.

The skip connection adds the feature maps before down-sampling to the feature maps before up-sampling by bypassing as shown in Fig. 1.

4To stabilize the training, we clip the gradient norm at 1.0 [32] and take an exponential moving average (EMA) of the network output [15] as,

$$\tilde{f}_{\theta^{(n,m+1)}}(z) = \eta \tilde{f}_{\theta^{(n,m)}}(z) + (1-\eta) f_{\theta^{(n,m+1)}}(z), \quad (21)$$
$$f_{\theta^{(n,0)}}(z) = \tilde{f}_{\theta^{(n,0)}}(z) = f_{\theta^{(n)}}(z), \quad (22)$$
$$f_{\theta^{(n+1)}}(z) = \tilde{f}_{\theta^{(n,M_2)}}(z), \quad (23)$$

where $\eta$ is a smoothing constant, $\tilde{f}$ is an EMA of DIP outputs, $m$ and $M_2$ are the number of sub-iteration and total sub-iteration of DIP for solving subproblem (13), respectively. In this study, we set $\eta = 0.99$.

*D. Overall algorithm*

The overall algorithm of the LM-DIPRecon method is shown in Algorithm 1.

---

**Algorithm 1** Algorithm of the LM-DIPRecon method

**First step**
1: Reconstruct the image with one main iteration of the LM-DRAMA.
2: Running Adam method with 1000 epochs to ready the initial network.
$$\theta^{(0)} = \underset{\theta}{\mathrm{argmin}} \left\| f_\theta(z) - x_{\mathrm{EM}}^{(1)} \right\|^2$$

**Second step**
Input: Iteration number $N$, Sub-iteration numbers $M_1$ and $M_2$, Initial network $\theta^{(0)}$, MR image $z$, positive constant $\rho$
1: $x^{(0)} = f_{\theta^{(0)}}(z)$
2: $\mu^{(0)} = \mathbf{0}$
3: **for** $n = 0$ to $N - 1$ **do**
4: $\quad x^{(n,0)} = x^{(n)}$
5: $\quad x_{\mathrm{base}} = f_{\theta^{(n)}}(z) - \mu^{(n)}$
6: $\quad$ **for** $m = 0$ to $M_1 - 1$ **do**
7: $\quad\quad u = nM_1 + m$
8: $\quad\quad k = \lceil u/N_{\mathrm{sub}} \rceil$ where $\lceil \cdot \rceil$ is round down.
9: $\quad\quad q = u \% N_{\mathrm{sub}}$
10: $\quad\quad \lambda = \frac{\beta}{\beta + q + \gamma k N_{\mathrm{sub}}}$
11: $\quad\quad x_{j,\mathrm{EM}} = x_j^{(n,m)} + x_j^{(n,m)} \lambda \left( \frac{N_{\mathrm{sub}}}{S_j} \sum_{t \in \mathrm{Sub}_q} \frac{a_{i(t)j}}{y_{i(t)}^{(n,m)}} - 1 \right)$
12: $\quad\quad x_j^{(n,m+1)} = \frac{x_{j,\mathrm{base}} - \frac{S_j}{\rho} + \sqrt{\left(x_{j,\mathrm{base}} - \frac{S_j}{\rho}\right)^2 + 4 x_{j,\mathrm{EM}} \frac{S_j}{\rho}}}{2}$
13: $\quad$ **end for**
14: $\quad x^{(n+1)} = x^{(n,M_1-1)}$
15: $\quad x_{\mathrm{label}} = x^{(n+1)} + \mu^{(n)}$
16: $\quad$ Running L-BFGS method with $M_2$ iterations to get
$$\theta^{(n+1)} = \underset{\theta}{\mathrm{argmin}} \| f_\theta(z) - x_{\mathrm{label}} \|^2$$
17: $\quad \mu^{(n+1)} = \mu^{(n)} + x^{(n,M_1-1)} - f_{\theta^{(n+1)}}(z)$
18: **end for**
19: **return** $\hat{x} = f_{\theta^{(N-1)}}(z)$

---

To ready the initial network $\theta^{(0)}$, we take a two-step approach [16]. At first, we reconstruct the image with one main-iteration of the LM-DRAMA and perform the MR-DIP as,

$$\theta^{(0)} = \underset{\theta}{\mathrm{argmin}} \left\| f_\theta(z) - x_{\mathrm{EM}}^{(1)} \right\|^2. \quad (24)$$

In this step, we update the network parameters by the Adam method [33] with 1000 epochs. We set one main-iteration of LM-DRAMA and 1000 epochs of MR-DIP as a relatively small number of iterations in the warmup of CNN because an initial image is preferred to be smooth.

In the second step, we iterate the three steps of the ADMM as shown in Algorithm 1. In each step of the ADMM, there is no need to solve the subproblems until convergence. For subproblem (12), we set $M_1 = 2$ and perform two sub-iterations of the regularized LM-DRAMA. For subproblem (13), we set $M_2 = 10$ and update the network parameters using the limited memory Broyden-Fletcher-Goldfarb-Shanno (L-BFGS) method with 10 epochs and a learning rate of 1.0. As the L-BFGS method has a fast convergence rate when the parameters are near a local optimum, we use the Adam method in the first step and the L-BFGS method in the second step. We followed the previous study [16] for the settings of $M_1$ and $M_2$

Last, after $N$ iterations, the algorithm outputs the image as,

$$\hat{x} = f_{\theta^{(N-1)}}(z), \quad (25)$$

where $\hat{x}$ is a final image.

## IV. EXPERIMENTAL SETUP

We evaluated the proposed methods using the simulation data and clinical data. We compared the proposed method relative to the LM-DRAMA and the MR-DIP. Image size was 128×128×80 and voxel size was 2.6 mm × 2.6 mm × 2.4 mm. The details of the data correction can be seen in the Appendix.

*A. Simulation data*

We performed Monte-Carlo simulation using our simulator. The simulation included the attenuation and scatter effects based on the delta-scattering method [34]. The positron range, angular deviation and random events were not included in the simulation. We made a digital brain phantom from a segmented MR image downloaded from the BrainWeb [35]. The digital brain phantom was set to a contrast of 1.00:0.25:0.05 for gray matter (GM), white matter (WM) and CSF. In addition, we inserted the three tumor regions with a contrast of 1.0, 1.1, and 1.2. Attenuation coefficients were set to 0.00958 mm$^{-1}$ for the soft tissue and CSF, and 0.0151 mm$^{-1}$ for the bone tissue.

The detector arrangement was the same as our brain PET scanner (Hamamatsu HITS-655000 [36]) with four-layer DOI. The detector unit consists of 32 × 32 lutetium-yttrium oxyorthosillicate (LYSO) scintillator array with a 1.2 mm pitch and 64 multi-pixel photon counters (MPPCs) in an 8 × 8 arrangement. To ensure that the load for the count rate in every layer is equally sensitive for gamma rays, the scintillator thickness was designed to be 3 mm, 4 mm, 5 mm, and 8 mm toward the outside of the field-of-view [36]. The scanner consists of 32 (radial) × 5 (axial) = 160 detector units. The number of crystals is $655 \times 10^3$, and the number of LORs is $107 \times 10^9$ including DOI bins.

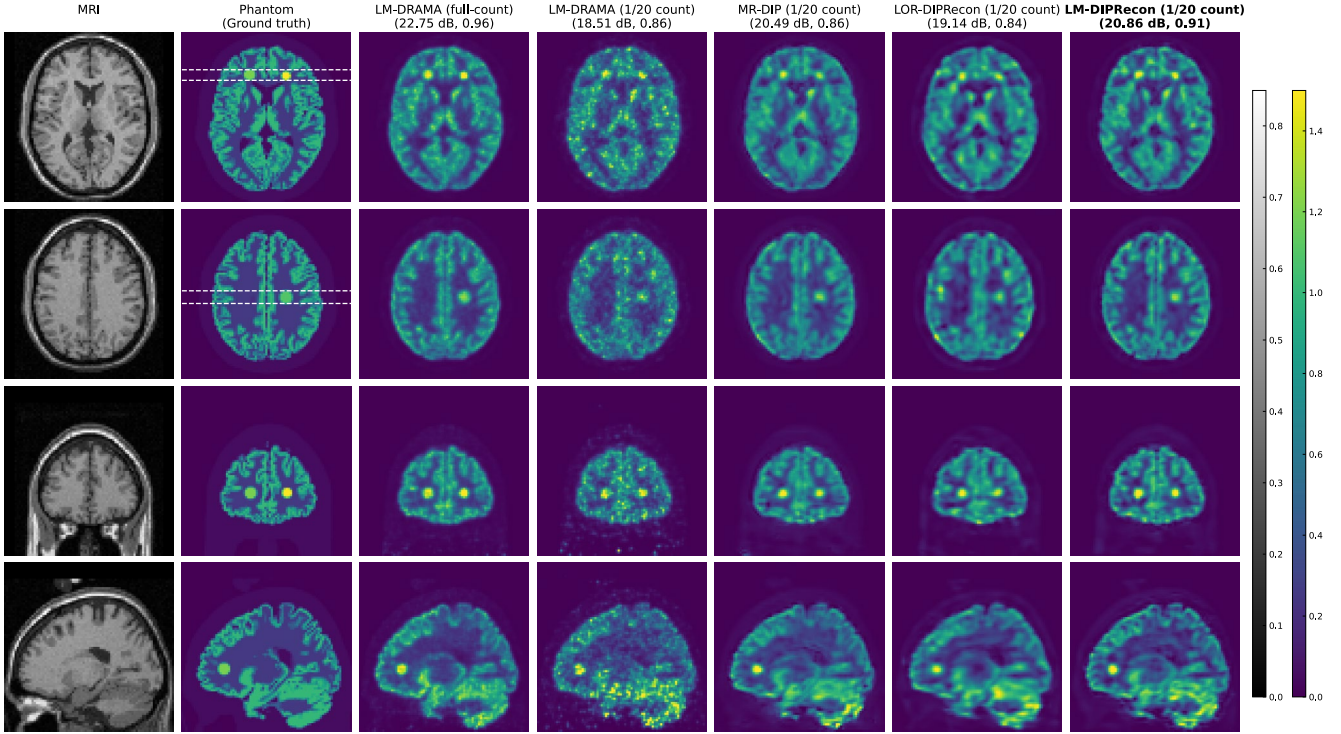

Fig. 2. Images of simulation data reconstructed by the proposed method and the other methods. From left to right, MRI and phantom image, LM-DRAMA with full-count, LM-DRAMA, MR-DIP, LOR-DIPRecon and LM-DIPRecon with 1/20 count. Each image is tagged with its PSNR and TR. White dot lines indicate the position of profiles in Fig. 3.

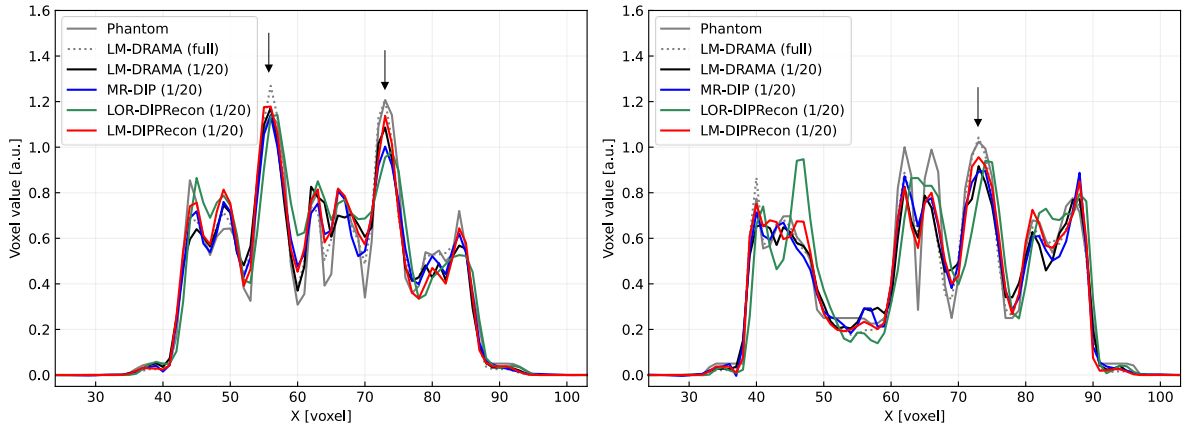

Fig. 3. Profiles on images of simulation data reconstructed by the proposed method and the other method. Left is a profile with six-pixel width across two small tumors, as indicated by the white dot lines on the first row of Fig. 2. Right is a profile with seven-pixel width across one large tumor, as indicated by the white dot lines on the second row of Fig. 2. Arrows indicate the position of the tumors.

The simulation included attenuation and scatter. We set the energy window of 400-650 keV with an energy resolution of 15%. The list data had $1.52 \times 10^8$ events. To evaluate the denoising performance, we thinned the list data into 1/20, $7 \times 10^6$ events. The number of LORs was $1.5 \times 10^4$ times the number of events.

The MR image was interpolated to the same dimension of the PET image and used as the prior distribution image of the DIP.

*B. Clinical data*

A clinical study was performed at Hamamatsu University School of Medicine. The Ethics Committees of Hamamatsu University School of Medicine approved the study, and written informed consent was obtained from the participant prior to enrollment. A healthy volunteer was scanned using HITS-655000 for 62 min after injection of 5 MBq/kg of $^{11}$C-MeQAA, a tracer for α7 nicotinic acetylcholine receptors which exist abundantly in the thalamus and striatum [37].

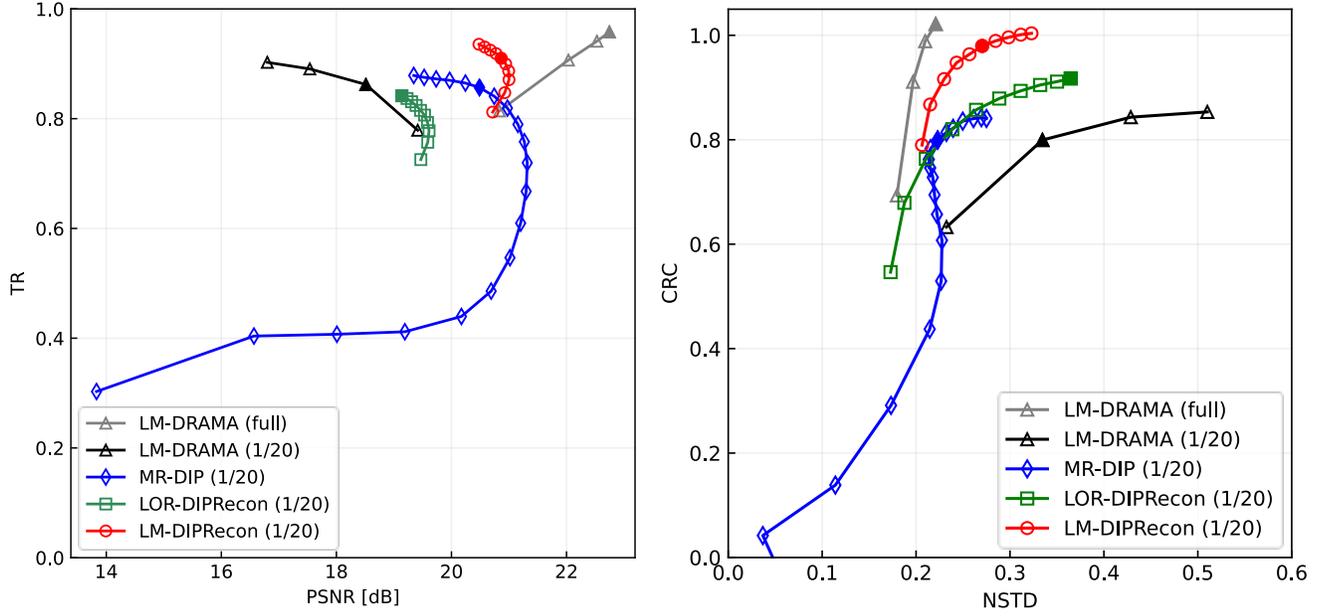

Fig. 4. Tradeoff curves between contrast and noise on the simulation study: TR versus PSNR (left) and CRC versus NSTD (right). Fill markers correspond to the images, as shown in Fig. 2. TR was calculated for the union of three tumor regions. CRC was calculated for the ROIs on the gray matter and tumor regions. NSTD was calculated for the ROIs on the white matter. The plots of LM-DRAMA correspond to 1, 2, 3, and 4 main-iterations. The plots of MR-DIP correspond to 100, 200, …, 2000 epochs. The plots of LOR-DIPRecon correspond to 32, 64,…, 320 iterations. The plots of LM-DIPRecon correspond to 20, 40,…, 200 iterations.

We separated the list data into three frames of 0-20 min, 20-42 min, and 42-62 min. We used the 42-62 min frame to evaluate the uptake of the QAA in the thalamus. The list data of 42-62 min had $4.7 \times 10^7$ events. Furthermore, to evaluate the denoising performance, we thinned the list data into 1/10, $4.7 \times 10^6$ events. The number of LORs was $2.3 \times 10^4$ times the number of events.

The MRI scan was performed using a 1.5 T MR scanner (Signa HDxt, GE, USA) with the following acquisition parameters: 3D mode sampling, Repetition Time/Time to Echo (TR/TE) (25/Minimum, T1-weighted), 30º flip angle, 1.5 mm slice thickness with no gap, and 256×256 matrices. The MR image was interpolated and aligned to the PET image using PMOD software (version 3.0, PMOD Technologies Ltd., Zurich, Switzerland) [38].

*C. Evaluation metrics*

In the simulation study, we evaluated the peak signal-to-noise ratio (PSNR) for quantitative evaluation.

$$\text{PSNR} = 10 \log_{10} \frac{\max(x_{\text{ref}})^2}{\frac{1}{N_R}\sum_{j \in R}(x_j - x_{j,\text{ref}})^2}, \quad (26)$$

where $x_{\text{ref}}$ is a phantom image, $R$ is a region of the whole brain, and $N_R$ is a number of voxels inside the whole brain.

For tumor quantification, the tumor value ratio (TR) was calculated as

$$\text{TR} = \frac{1}{N_{\text{tumor}}} \sum_{j \in R_{\text{tumor}}} \frac{x_j}{x_{j,\text{ref}}}, \quad (27)$$

where $R_{\text{tumor}}$ is a union of the three tumor regions and $N_{\text{tumor}}$ is a number of voxels inside the tumor regions.

In addition, we evaluated the curves of contrast recovery coefficient (CRC) versus normalized standard deviation (NSTD) to evaluate the tradeoff between contrast and noise. The CRC was calculated by the following equations.

$$\text{CRC} = \left(\frac{\bar{a}}{\bar{b}} - 1\right) \bigg/ \left(\frac{\bar{a}_{\text{ref}}}{\bar{b}_{\text{ref}}} - 1\right), \quad (28)$$

$$\bar{a} = \frac{1}{K_a} \sum_{k=1}^{K_a} a_k, \quad (29)$$

$$\bar{b} = \frac{1}{K_b} \sum_{k=1}^{K_b} b_k, \quad (30)$$

where $\bar{a}$ is the average of multiple region of interest (ROI) uptakes set to GM and tumor, $\bar{b}$ is the average of multiple ROI values set to WM, $K_a$ is the number of ROIs on the GM and tumor, $K_b$ is the number of ROIs on WM, $a$ is an ROI uptake on the GM and tumor, and $b$ is an ROI value on the WM.

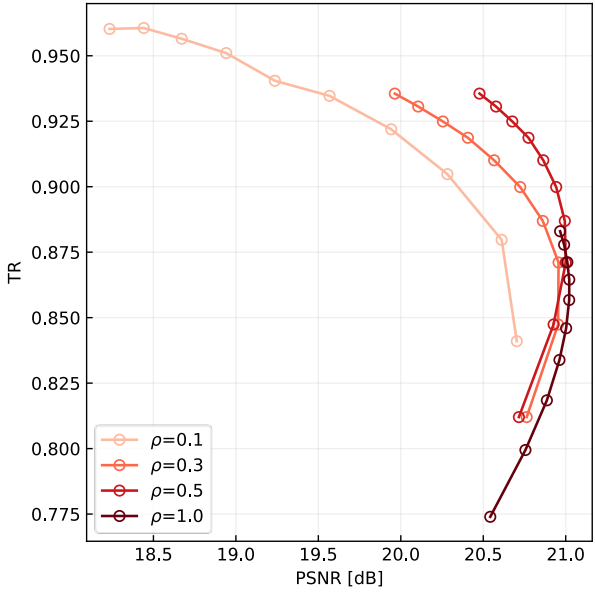

Fig. 5. Effects of hyper parameter $\rho$ to the tradeoff curves between TR and PSNR in the simulation study with 1/20 count.

$$a_k = \frac{1}{N_{a,k}} \sum_{j \in R_{a,k}} x_j, \quad (31)$$

$$b_k = \frac{1}{N_{b,k}} \sum_{j \in R_{b,k}} x_j, \quad (32)$$

Where $N_{a,k}$ is the number of voxels inside the $k$-th ROI on the GM and tumor, $N_{b,k}$ is the number of voxels inside the $k$-th ROI on the WM, $R_{a,k}$ is the $k$-th ROI on the GM and tumor, and $R_{b,k}$ is the $k$-th ROI on the WM.

The NSTD can be calculated by the following equation.

$$\text{NSTD} = \frac{1}{\bar{b}} \sqrt{\frac{1}{K_b} \sum_{k=1}^{K_b} (b_k - \bar{b})^2}. \quad (33)$$

In this study, $K_a = 20$ for the GM. $K_a = 18$ for the tumor, and $K_b = 25$ for the WM.

In the clinical study, for ROI uptake quantification, the uptakes were calculated for the ROI on the thalamus. In addition, STD was calculated for the reconstructed images of clinical data.

In both simulation and clinical studies, we compared the following methods and parameters.

- **LM-DRAMA:** We performed the LM-DRAMA with 1, 2, 3, and 4 main-iterations. We applied the 3D Gaussian filter with full-width-at-half-maximum of 3 mm as a post smoothing. For display, we chose the LM-DRAMA with two main-iterations because it is a default setting of the HITS-655000.
- **MR-DIP:** We performed the MR-DIP using the reconstructed image of LM-DRAMA with two main-iteration, as the label. We used the Adam method and performed 2000 epochs at maximum.
- **LOR-DIPRecon**: We performed the sinogram-based DIPRecon [16] as LOR-DIPRecon. The events were accumulated in the LOR sinogram which is a sinogram not interpolated to evenly space and maintains a raw sampling pitch of LOR in a radial direction. To reduce the number of LORs, the deep pairs of DOI layer were combined to shallow pairs of DOI layer by the DOI compression method [39]. We compressed the 16 DOI pairs to one pair. In addition, we bundled the crystals into $4 \times 4$ in radial and axial directions to reduce the number of LORs. The ring difference was sampled with the maximum ring difference of ±40 and span of three. In the results, the size of the LOR sinogram was 128 (radial) × 128 (azimuth) × 80 (slice) × 27 (ring difference). The image reconstruction from LOR sinograms was performed by the LOR-DRAMA which is the LOR-OSEM method [40] with the subset-dependent relaxation [23]. We set the $N_{\text{sub}} = 32$ and $\beta = 50$ for the LOR-DRAMA. For the LOR-DIPRecon, we performed the two-step approach. In the first step, we performed the MR-DIP using Adam method with 1000 epochs, where the label was the reconstructed image by one main-iteration of LOR-DRAMA. The trained network by MR-DIP was used as the initial network parameter $\theta^{(0)}$. In the second step, we performed 320 iterations of LOR-DIPRecon at maximum. The other hyper parameters were the same as the LM-DIPRecon.
- **LM-DIPRecon:** We performed the two-step approach as explained above. In the first step, we performed the MR-DIP using the Adam method with 1000 epochs, where the label was the reconstructed image by one main-iteration of LM-DRAMA. The trained network by MR-DIP was used as the initial network parameter $\theta^{(0)}$. In the second step, we performed 200 iterations of Algorithm 1 at maximum with sub-iterations of $M_1 = 2$ and $M_2 = 10$ for subproblem (12) and (13), respectively. We used the L-BFGS method for subproblem (13) and set $\rho = 0.5$ for the regularization.

## V. RESULTS

Fig. 2 shows the images of simulation data reconstructed by the proposed method and the other methods. The images were tagged with their PSNRs and TRs. The reconstructed image of LM-DRAMA with 1/20 count was noisy because the post smoothing was limited to 3 mm. Applying MR-DIP reduced the noise at a slight cost of the contrast. The LM-DIPRecon method maintained the contrast and provided sharper images than those of MR-DIP and LOR-DIPRecon.

Fig. 3 shows the profiles across the tumor regions. The peaks of profile on MR-DIP and LOR-DIPRecon at tumor positions tended to be lower than the phantom as shown in Fig. 3. The LM-DIPRecon obtained closer peaks to the phantom than the

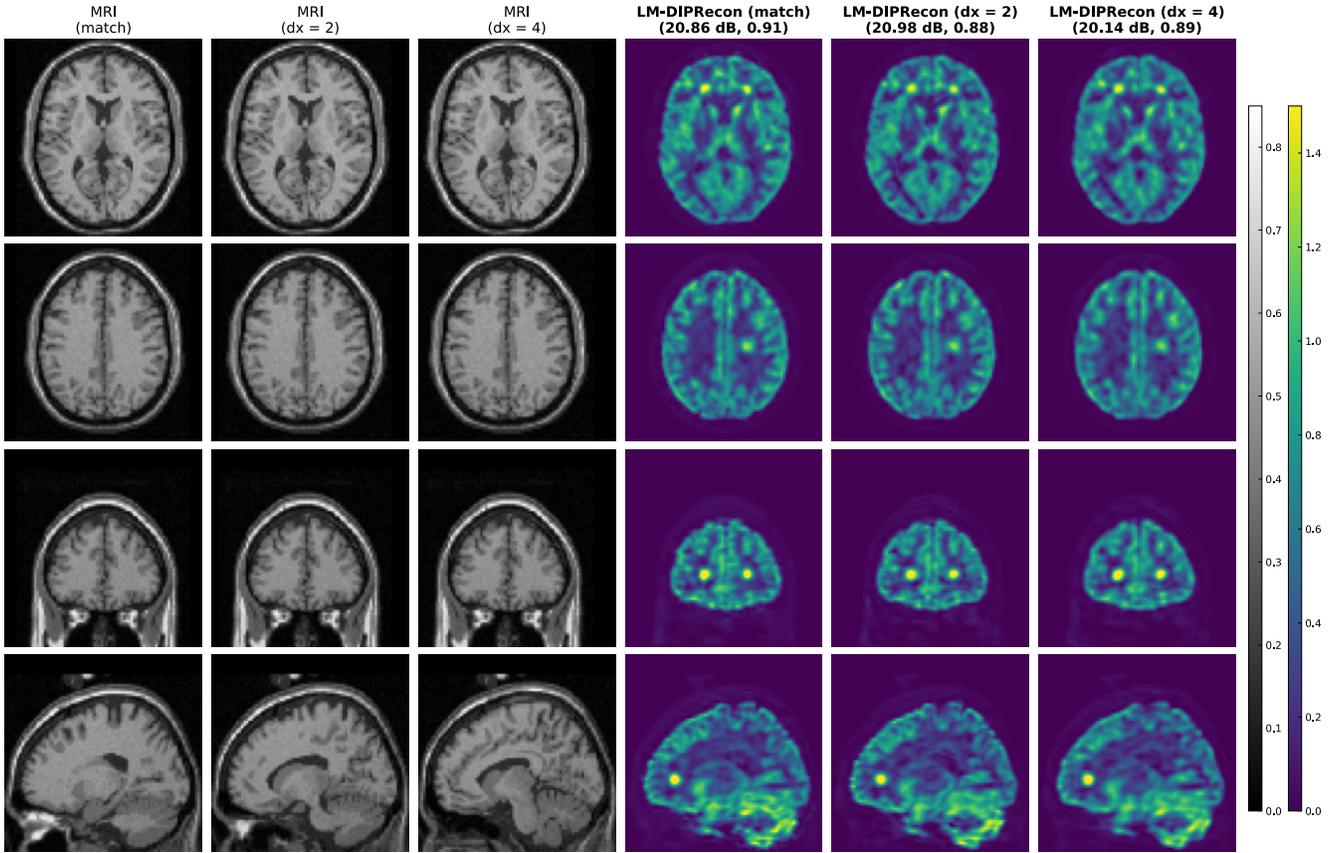

Fig. 6. Images of simulation data with 1/20 count when there is misregistration between MR and PET images. From left to right, MRIs with zero, two and four voxel shift in $x$ direction, images of LM-DIPRecon using MRIs with zero, two and four voxel shift in $x$ direction. The images of LM-DIPRecon are tagged with TR and PSNR. $dx$ is the amount of shift in voxel.

MR-DIP and LOR-DIPRecon.

Fig. 4 shows the tradeoff curves between contrast and noise in the simulation study: TR vs. PSNR and CRC vs. NSTD. For the LM-DRAMA with 1/20 count, the noise increased as the main-iteration increased. The MR-DIP improved the PSNR and NSTD relative to the LM-DRAMA with 1/20 count when the TR and CRC were the same as those of LM-DRAMA with 1/20 count. The LM-DIPRecon maintained the same contrast as LM-DRAMA with full-count while improving PSNR and NSTD relative to the LM-DRAMA with 1/20 count. The LM-DIPRecon provided the better PSNR than that of LOR-DIPRecon. In addition, it showed the better CRC than that of LOR-DIPRecon in the case of the same NSTD.

To assess how sensitive the proposed method is to hyperparameters, we performed the LM-DIPRecon with $\rho$ of 0.1, 0.3, 0.5 and 1.0 on the simulation data with 1/20 count. Among the hyperparameters, primarily $\rho$ determines the tradeoff between contrast and noise by controlling the strength of regularization by MR-DIP. Fig. 5 shows the tradeoff curves between TR and PSNR for different values of $\rho$. $\rho = 0.5$ provided the best tradeoff curve.

To assess how sensitive the proposed method is to mismatch between MR and PET images, we deliberately shifted the MR image in the $x$ direction with two or four voxels in simulation data, and performed the LM-DIPRecon. Fig. 6 and 7 shows the images and the tradeoff curves between contrast and noise, respectively, on the simulation data with 1/20 count when the misregistration between the MR and PET images was zero, two, and four voxels in $x$ direction. When the shift of MRI in the $x$ direction was four voxels, the PSNR decreased about 0.7 dB on average.

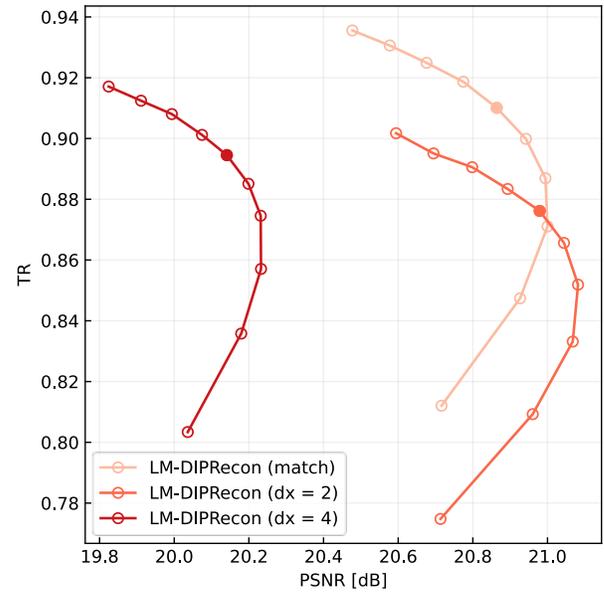

Fig. 7. Tradeoff curves between contrast and noise on the simulation study with 1/20 count when the misregistration between MR and PET images is zero, two, and four voxel in $x$ direction. $dx$ is the amount of shift in voxel.

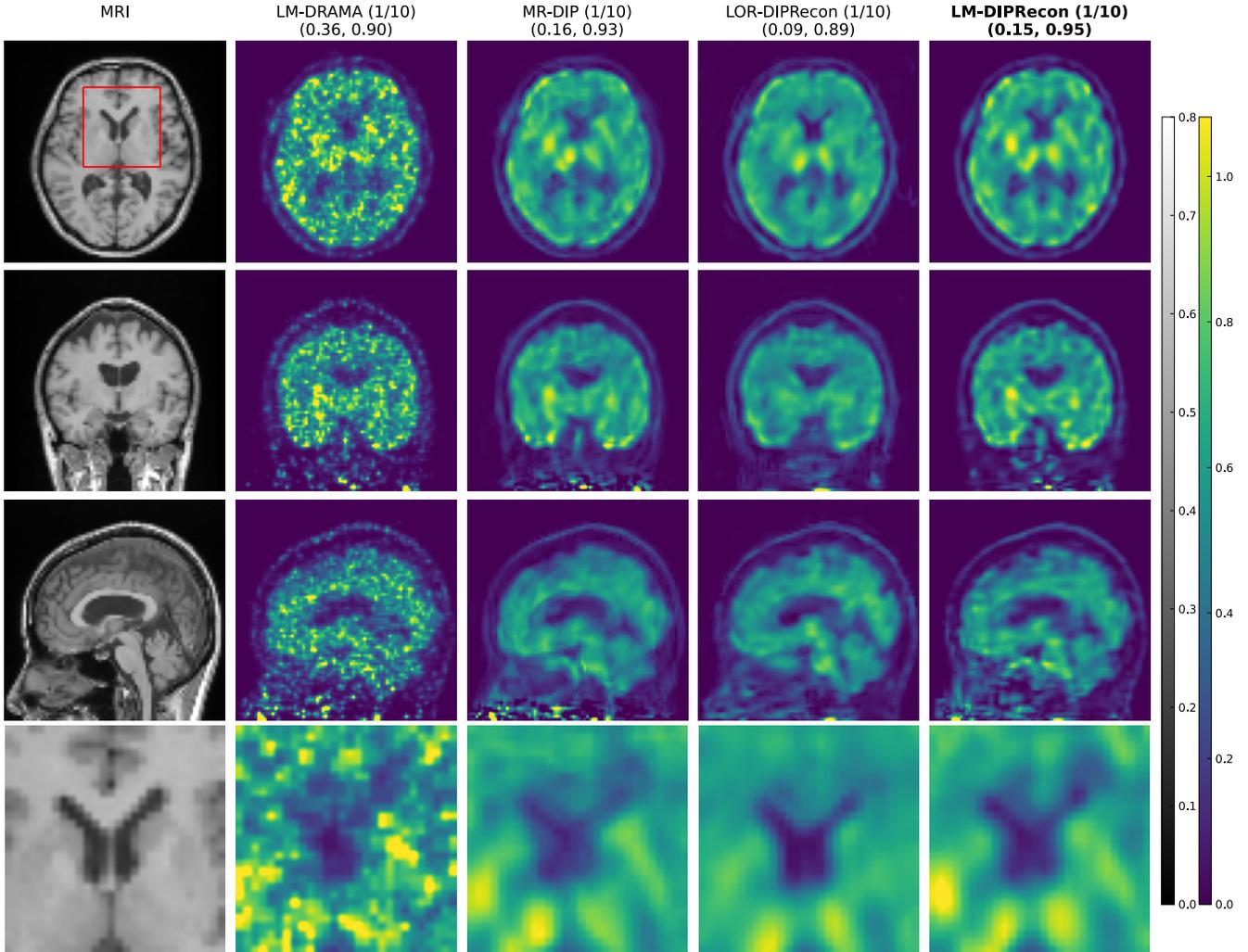

Fig. 8. Images of clinical data with 1/10 count reconstructed by the proposed method and the other methods. From left to right, MRI, LM-DRAMA, MR-DIP, LOR-DIPRecon and LM-DIPRecon. The images are tagged with the NSTD and the uptake. The fourth row shows the zoom-in images on the red square region.

Fig. 8 shows the images of clinical data with 1/10 count reconstructed by the proposed method and the other methods. The images showed that the $^{11}$C-MeQAA accumulated dominantly in the thalamus and striatum [37]. Although the reconstructed images at 1/10 count were very noisy, MR-DIP, LOR-DIP and LM-DIPRecon reduced the noises. In addition, LM-DIPRecon restored the gray matter structure more clearly than MR-DIP. The LOR-DIPRecon provided the smoother image than that of LM-DIPRecon.

Fig. 9 shows the tradeoff curves between contrast and noise in the clinical study with 1/10 count: Uptake vs. NSTD. The LM-DIPRecon provided a higher uptake than those of the other methods.

Fig. 10 shows the images of clinical data with full-count reconstructed by the proposed method and the other methods. The LM-DIPRecon imaged the clearer structure of gray matter than the other methods and reduced the noise relative to LM-DRAMA.

## VI. DISCUSSION

In this study, we incorporated DIP into list-mode PET image reconstruction using an ADMM framework. As the computation efficiency of list-mode PET image reconstruction depends on the number of events and not on the number of LORs, the proposed LM-DIPRecon method will be suitable for a PET scanner with many LORs and additional information such as TOF and DOI.

In the simulation study, LM-DIPRecon provided sharper images visually than those of MR-DIP (Fig 2). This is because the number of iterations of list-mode PET image reconstruction in LM-DIPRecon is virtually larger than that of MR-DIP. The MR-DIP train a CNN to map an MR image to a PET image reconstructed using LM-DRAMA with two main-iterations. Therefore, the degree of convergence of the MR-DIP image to the maximum likelihood solution is limited to the level of two main-iterations. On the other hand, LM-DIPRecon performs more iterations of regularized LM-DRAMA through solving subproblem (10). For example, $N = 200$ and $M_1 = 2$ correspond to 10 main-iterations with 40 subsets. Hence, the LM-DIPRecon method could reconstruct a more convergent image than the MR-DIP within the constraint that the image is representable by the CNN. LM-DIPRecon started optimization with CNN prepared by the first step, but MR-DIP is also different in that it performed optimization from random initialization. This is a reason why the images of LM-DIPRecon were sharper than those of MR-DIP. In addition, LM-DIPRecon provided sharper image than that of LOR-DIPRecon because it enables full use of DOI information.

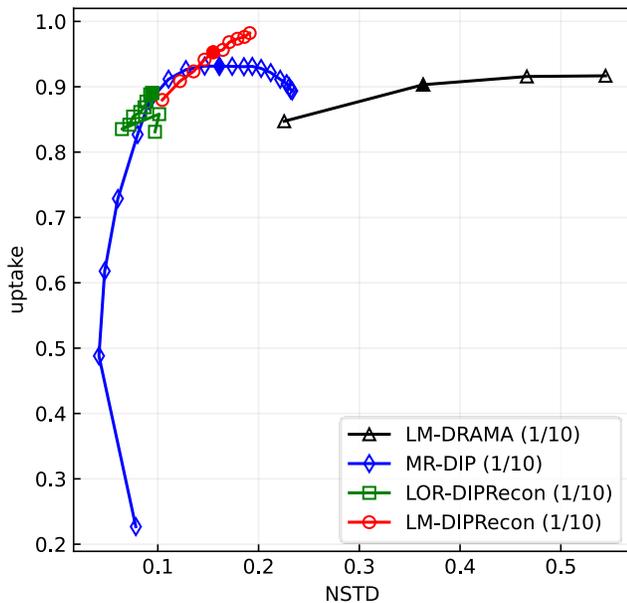

Fig. 9. Tradeoff curves between contrast and noise on the clinical study with 1/10 count: Uptake versus NSTD. Fill markers correspond to the images as shown in Fig. 5. Uptake was calculated for the ROI on the thalamus. NSTD was calculated for the ROIs on the white matter. The plots of LM-DRAMA correspond to 1, 2, 3, and 4 main-iterations. The plots of MR-DIP correspond to 100, 200, …, 2000 epochs. The plots of LOR-DIPRecon correspond to 32, 64,…, 320 iterations. The plots of LM-DIPRecon correspond to 20, 40, …, 200 iterations.

The LM-DIPRecon method provided better tradeoff curves between contrast and noise than those of the other methods (Fig. 4). However, in some cases, the tradeoff curves of LM-DIPRecon and MR-DIP overlapped (Fig 4. Left). When the computation time is prioritized, MR-DIP may become the preferred option instead of LM-DIPRecon. However, the merit of LM-DIPRecon is consistency with measured data. The MR-DIP method can be consistent with a label image, but the consistency with the measured data is limited by the label image in principle. Hence, the risk of eliminating the signal such as tumor by the non-linear image processing such as MR-DIP may be reduced by the LM-DIPRecon. When the data consistency is prioritized, the LM-DIPRecon is recommended. The LM-DIPRecon provided better contrast than that of LOR-DIPRecon in the case of the same noise level.. This is because the list-mode approaches can fully use the DOI information.

The LM-DIPRecon provided the close tradeoff curves when $\rho$ is within 0.3 to 1.0 (Fig. 5). When $\rho$ is 0.1, the strength of regularization was not sufficient to suppress the noise. We expect that the proposed method is stable for the choice of $\rho$ within 0.3 to 1.0.

The TR seemed more robust to misregistration between MRI and PET than PSNR (Fig. 7). This is because the tumors were not included in MRI from the beginning, and TR was not affected by the misregistration. The PSNR was relatively sensitive to misregistration between MRI and PET (See the tradeoff curve with $dx = 4$ in Fig. 7). This is likely because the misregistration of piecewise constant regions with large volume such as gray matter and white matter weaken the noise suppression performance of MR-DIP. From these results, although the proposed method is considered to be robust to the absence of signals such as tumors in MRI, it is considered to be relatively sensitive to misregistration between MRI and PET in terms of noise.

In the simulation study, a tumor on the image of the proposed method seemed to have higher contrast than the ground truth (Fig. 2). We could consider that this is also because of the mismatch between MRI and PET. We can consider that the noise suppression by the MR-DIP is weak on the tumor regions because the tumors were not included in the MR images. In addition, it was reported that the maximum likelihood (ML) reconstructions cause edge and noise artifacts as progress to convergence [41]. We deduce that ML reconstruction-caused artifacts on tumor regions could not be suppressed by MR-DIP because of tumor absence in MRI. These two factors are considered to have caused the overshoot of tumor uptake.

In the clinical study with 1/10 count, LOR-DIPRecon provided a smoother image than that of LM-DIPRecon (Fig. 8). This can be considered as the smoothing effect of the DOI compression and crystal bundling. However, this smoothing may weaken the contrast. For example, the fourth row of Fig. 10 zoomed in the red square region in the first row of Fig. 10, where these are the results of clinical study with full-count. The red arrow indicated the hot spot which can be seen by the LM-DIPRecon, but not by the LOR-DIPRecon. This indicated that the smoothing effect by the DOI compression and crystal bundling has a risk to obscure the hot spot. In contrast, the LM-DIPRecon may be helpful to find the hot spot. In the clinical study, LM-DIPRecon visualized a clearer structure of GM than MR-DIP (Fig. 8 and 10). The image quality of real data depends on factors such as an imperfection of data correction or a subject motion. Note that these factors degrade the image quality and may make the visual difference between the reconstruction methods small. The LM-DIPRecon method provided a better tradeoff curve of uptake vs. NSTD than the other methods, similar to the simulation study (Fig. 9).

Currently, the number of LORs of research tomographs have achieved the order of $10^9$. For example, the number of LORs of uEXPLORER total body PET/CT scanner is $92 \times 10^9$ [42]. In addition, the number of LORs of brain PET scanner with four-layer DOI detectors which was used in this study is $107 \times 10^9$

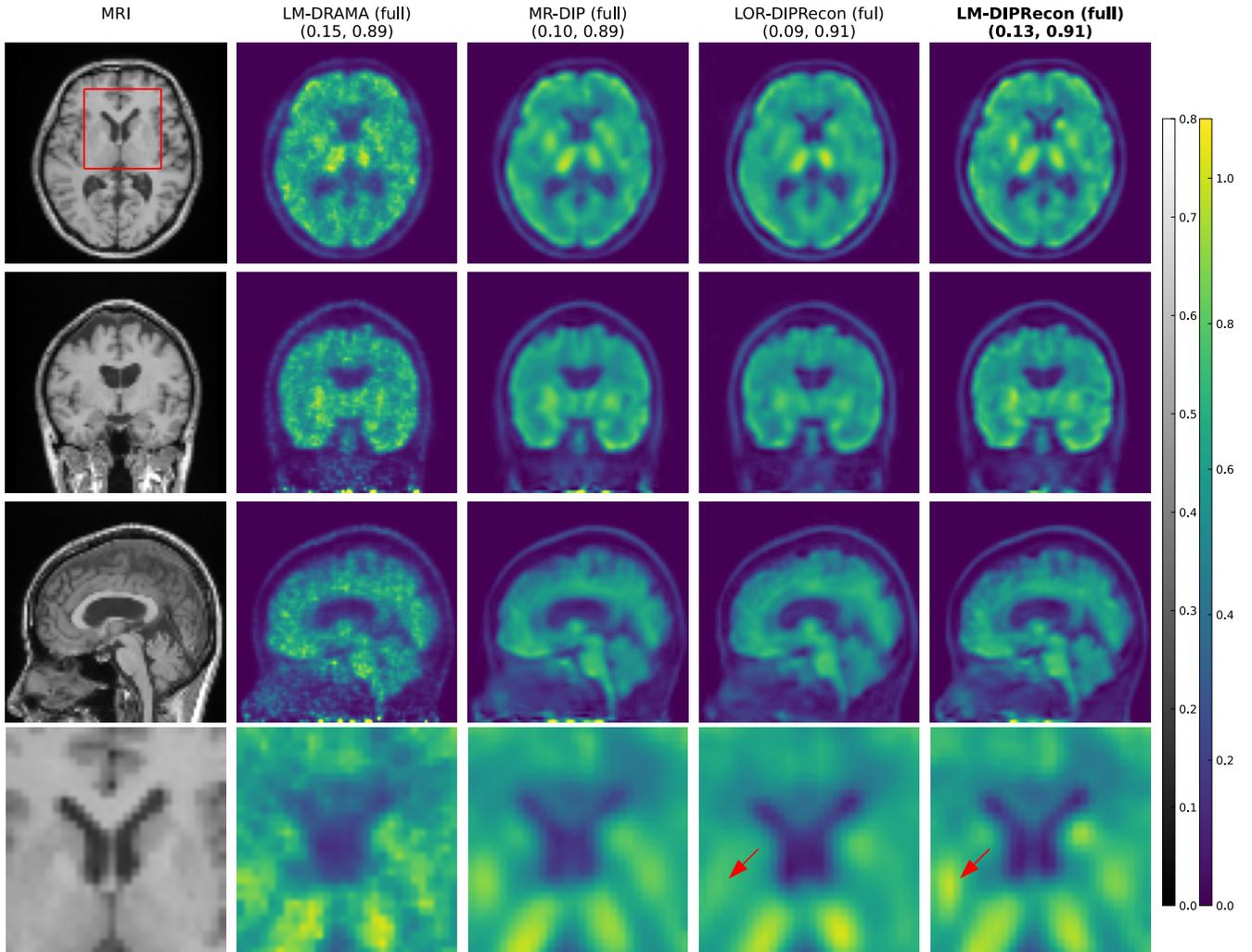

Fig. 10. Images of clinical data with full-count reconstructed by the proposed method and the other methods. From left to right, MRI, LM-DRAMA, MR-DIP, LOR-DIPRecon and LM-DIPRecon. The images are tagged with the NSTD and the uptake. The fourth row shows the zoom-in images on the red square region. The red arrow indicates a convincing accumulation of $^{11}$C-MeQAA in the striatum on the LM-DIPRecon, but not on the LOR-DIPRecon.

[36]. To full-use of the measured information in these scanners, the list-mode PET image reconstruction is still needed. The simulation and clinical results of the brain PET scanner with DOI detectors demonstrated the feasibility of the LM-DIPRecon for the PET scanners with a large number of LORs.

The limitation of the LM-DIPRecon is a long computation time. In this study, the regularized LM-DRAMA was computed using six cores of Intel i7-5930k 3.5GHz through the message passing interface, and the MR-DIP was computed using NVIDIA Quadro P6000 graphics card. In this condition, the two sub-iterations of regularized LM-DRAMA took 2-3 seconds, and the 10 sub-iterations of L-BFGS method for the MR-DIP took about 96 seconds. Hence, 200 iterations of LM-DIPRecon took about $2\times10^4$ seconds in our environment. As the computational cost of MR-DIP was higher than regularized LM-DRAMA, we could reduce the total computational cost by increasing the number of sub-iterations of the regularized LM-DRAMA and decreasing the number of iterations of the LM-DIPRecon.

Another limitation of this study is a mismatch between MR and PET images. In simulation data, the structures of MR and PET images had agreed except for tumor regions, but not for the clinical data. This is because MRI and PET were taken separately in this study. This mismatch can be considered as one reason why the performance of the proposed method degraded in clinical data compared to simulation data. We expect that the MR-PET scanner easily eliminates this problem.

In this study, we demonstrated the proposed method on the four-layer DOI-PET scanner [36]. Another demonstration of the proposed method on TOF-PET scanners is also an important future study, and it is expected to strongly support the usefulness of the proposed method. In near future, we plan to evaluate the proposed method using clinical data of the dedicated brain TOF-PET scanner [43].

In the future, an extension of LM-DIPRecon to 4D image reconstruction [44], [45] can be considered. As list-mode data has finer temporal information than dynamic sinograms, LM-4D-DIPRecon may be useful for dynamic PET imaging or direct parametric imaging [46]. For example, 4D image reconstruction models the time activity curves as a linear combination of temporal basis functions, and estimates the coefficients of linear combination directly from dynamic sinogram or list data [44], [47]. By choosing temporal basis functions with correlation, noise can be suppressed by 4D PET image reconstruction. For list data, we can compute the contribution of event to the temporal basis function in millisecond order, but in dynamic sinogram we restricted to frame time unit and, which may result in a loss of some information. Hence, the proposed method is also promising as a framework for 4D PET image reconstruction using DIP [17]. The high temporal resolution of list data is also useful for motion correction [48], [49]. From these ideas, the LM-DIPRecon can be expected to be useful for the 4D and 5D problems involving dynamic imaging of moving organs [50]. In this study, the MR image was used as a prior distribution image of the DIP, but an MR-guided deep decoder (MR-GDD) [51], which uses the MR image through an attention mechanism, has already been proposed. As the MR-GDD works to improve the contrast of the tumor which does not exist on the MR image, the list-mode PET image reconstruction using MR-GDD may improve the contrast of the tumor relative to the LM-DIPRecon.

## VII. Conclusion

We proposed list-mode PET image reconstruction using DIP which is the first trial to integrate list-mode PET image reconstruction and CNN. We incorporated DIP into list-mode PET image reconstruction using an ADMM framework. We evaluated the proposed method using simulation and clinical data. The proposed method provided a sharper image, higher contrast, and lower noise than the other methods. These results indicated that the proposed method is promising for the next generation of brain PET scanners.

## Appendix

In the experiments, data corrections [52] and the shift-invariant image-space point-spread-function (PSF) [53] were incorporated in the LM-DRAMA as,

$$\tilde{x}_{j,\text{EM}}^{(k,q+1)} = x_j^{(k,q)} +$$
$$x_j^{(k,q)} \lambda^{(k,q)} \left( \frac{N_{\text{sub}}}{\tilde{S}_j} \sum_{j'} P_{jj'} \sum_{t \in \text{Sub}_q} \frac{a_{i(t)j'} \delta(t)}{\tilde{y}_{i(t)}^{(k,q)}} - 1 \right), \quad (34)$$

$$x_{j,\text{EM}}^{(k,q+1)} = \max\left(\tilde{x}_{j,\text{EM}}^{(k,q+1)}, 0\right), \quad (35)$$

$$\tilde{y}_i^{(k,q)} = A_i \left( \sum_j a_{ij} \sum_{j'} P_{jj'} x_{j'}^{(k,q)} \right) + \frac{C_i}{B_i}, \quad (36)$$

$$\tilde{S}_j = \sum_j P_{jj'} \sum_i B_i a_{ij'}, \quad (37)$$

$$A_i = e^{-\sum_j a_{ij} \mu_j}, \quad (38)$$

$$\delta(t) = \begin{cases} 1 & t \text{ is prompt} \\ -1 & t \text{ is delayed} \end{cases}, \quad (39)$$

where $A$ is an attenuation probability, $B$ is an efficiency of LOR, $C$ is a scatter component, $P$ is a PSF, $\tilde{S}$ is a sensitivity image including the LOR efficiency and the PSF, $\mu$ is an image of linear attenuation coefficient ($\mu$-map), and $\delta$ is a sign of event for the random correction of the delayed subtraction.

The efficiency of LOR was estimated by the component-based normalization technique [54]. The scatter component was estimated by the single scatter simulation method [55]. In the simulation study, the $\mu$-map of the phantom image was used for the attenuation correction (AC) and the scatter estimation. In the clinical study, the $\mu$-map was estimated by the segmentation of the reconstructed image without AC. The bone tissues were inserted in the surface of segmented image using dilation and erosion filters. The PSF was modeled by the 3-D Gaussian function with full-width at half-maximum of one voxel because the voxel size was larger than the spatial resolution of HITS-655000 [36].

## Acknowledgment

We thank the members of the fifth research group in the central research laboratory of the Hamamatsu Photonics K. K. for their kind support.

## References

[1] M. E. Phelps, *PET: Molecular Imaging and Its Biological Applications*. New York, NY, USA: Springer-Verlag, 2004. [Online]. Available: https://www.springer.com/la/book/9780387403595
[2] J. Qi *et al.*, "High-resolution 3D Bayesian image reconstruction using the microPET small-animal scanner," *Phys. Med. Biol.*, vol. 43, no. 4, pp. 1001-1014, 1998.
[3] J. Nuyts *et al.*, "A concave prior penalizing relative differences for maximum-a-posterior reconstruction in emission tomography," *IEEE Trans. Nucl. Sci.*, vol. 49, no. 1, pp. 56-60, 2002.
[4] G. Wang and J. Qi, "PET image reconstruction using kernel methods," *IEEE Trans. Med. Imaging*, vol 34, no. 1, pp. 61-71, 2015.
[5] B. Bai, Q. Li, R. M. Leahy, "Magnetic resonance-guided positron emission tomography image reconstruction," *Semin. Nucl. Med.*, vol.43, no. 1, pp. 30-44, 2013.
[6] C. Comtat *et al.*, "Clinically feasible reconstruction of 3D whole-body PET/CT data using blurred anatomical labels," *Phys. Med. Biol.*, vol. 47. No. 1, pp. 1-20, 2002.
[7] A. J. Reader et al., "Deep learning for PET image reconstruction," *IEEE Trans. Radiat. Plasma Med. Sci.*, vol. 5, no. 1, pp. 1-25, 2020.
[8] K. Gong *et al.*, "The evolution of image reconstruction in PET: From filtered back-projection to artificial intelligence," *PET Clin.*, vol. 16, no. 4, pp. 533-542, 2021.
[9] A. Mehranian and A. J. Reader, "Model-based deep learning PET image reconstruction using forward-backward splitting expectation-maximization," *IEEE Trans. Radiat. Plasma Med. Sci.*, vol. 5, no. 1, pp. 54-64, 2020.
[10] K. Ote and F. Hashimoto, "Deep-learning-based fast TOF-PET image reconstruction using direction information," *Radiol. Phys. Technol.*, vol. 15, pp. 72-82, 2022.
[11] B. Zhu et al., "Image reconstruction by domain-transform manifold learning," *Nature*, vol. 555, no. 7697, pp. 487-492, 2018.
[12] W. Whiteley, W. K. Luk, and J. Gregor, "DirectPET: full-size neural network PET reconstruction from sinogram data," *J. Med. Imaging*, vol. 7, no. 3, 032503, 2020.
[13] I. Häggström *et al.*, "DeepPET: A deep encoder–decoder network for directly solving the PET image reconstruction inverse problem," *Med. Image Anal.*, vol. 54, pp. 253–262, 2019.
[14] K. Gong *et al.*, "Iterative PET image reconstruction using convolutional neural network representation," *IEEE Trans. Med. Imaging*, vol. 38, no. 3, pp. 675-685, 2018.
[15] D. Ulyanov, A. Vedaldi and V. Lempitsky, "Deep image prior," *Int. J. Comput. Vis.*, vol. 128, pp. 1867-1888, 2020.

[16] K. Gong *et al.*, "PET image reconstruction using deep image prior," *IEEE Trans. Med. Imaging*, vol. 38, no. 7, pp. 1655-1665, 2018.

[17] T. Yokota *et al.*, "Dynamic PET image reconstruction using non-negative matrix factorization incorporated with deep image prior," In *2019 IEEE/CVF International Conference on Computer Vision (ICCV)*, Seoul, Korea, pp. 3126-3135, 2020.

[18] F. Hashimoto, K. Ote, and Y. Onishi, "PET image reconstruction incorporating deep image prior and a forward projection model," *IEEE Trans. Radiat. Plasma Med. Sci.*, Early Access, 2022. 10.1109/TRPMS.2022.3161569

[19] M. G, Spangler-Bickell *et al.*, "Ultra-fast list-mode reconstruction of short PET frames and example applications," *J. Nucl. Med.*, vol. 62, no. 2, pp. 287-292, 2021.

[20] S. Boyd *et al.*, "Distributed optimization and statistical learning via the alternating direction method of multipliers," *Found. Trends Mach. Learn.*, vol. 3, no. 1, pp. 1-122, 2011.

[21] X. Cao, Q. Xie, and P Xiao, "A regularized relaxed ordered subset list-mode reconstruction algorithm and its preliminary application to undersampling PET imaging," *Phys. Med. Biol.*, vol. 60, no. 1, pp. 49-66, 2015.

[22] T. Nakayam and H. Kudo, "Derivation and implementation of ordered-subsets algorithms for list-mode PET data," *2005 IEEE Nucl. Sci. Symp. Conf. Rec.*, Fajardo, PR, USA, pp. 1950-1954, 2005.

[23] E. Tanaka and H. Kudo, "Subset-dependent relaxation in block iterative algorithms for image reconstruction in emission tomography," *Phys. Med. Biol.*, vol. 48, no. 10, 1405-1422, 2003.

[24] J. Qi, "Calculation of the sensitivity image in list-mode reconstruction for PET," *IEEE Trans. Nucl. Sci.*, vol. 53, no. 5, pp. 2746-2751, 2006.

[25] J. Lehtinen *et al*, "Noise2Noise: Learning image restoration without clean data," *arXiv preprint arXiv:1803.04189*, 2018.

[26] J. Cui *et al.*, "PET image denoising using unsupervised deep learning," *Eur. J. Nucl. Med. Mol. Imaging*, vol. 46, no. 13, pp. 2780-2789, 2019.

[27] F. Hashimoto *et al.*, "Dynamic PET image denoising using deep convolutional neural networks without prior training datasets," *IEEE Access*, vol. 7, pp. 96594-96603, 2019.

[28] K. Lange, D. R. Hunter, and I. Yang, "Optimization transfer using surrogate objective functions," *J. Comput. Graph. Stat.*, vol. 9, no. 1, pp. 1-20, 2000.

[29] G. Wang and J. Qi, "Penalized likelihood PET image reconstruction using patch-based edge-preserving regularization," *IEEE Trans. Med. Imaging*, vol. 31, no. 12, pp. 2194-2204, 2012.

[30] E. S. Helou Neto and A. R. De Pierro, "Convergence results for scaled gradient algorithms in positron emission tomography," *Inverse Problems*, vol. 21, no. 6, pp. 1905-1914, 2005.

[31] Ö. Çiçek *et al.*, "3D U-Net: learning dense volumetric segmentation from sparse annotation," *Medical Image Computing and Computer Assisted Intervention* (MICCAI), LNCS, vol. 9901, pp. 424–432, 2016.

[32] R. Pascanu, T. Mikolov, and Y. Bengio, "On the difficulty of training recurrent neural networks," *arXiv preprint arXiv: 1211.5063*, 2013

[33] D. P. Kingma and J. Ba, "Adam: a method of stochastic optimization," *arXiv preprint arXiv:1412.6980*, 2017

[34] M. Ljungberg A. Larsson, L. Johansson, "A new collimator simulation in SIMIND based on the delta-scattering technique," *IEEE Trasns. Nucl. Sci.*, vol. 52, no. 5, pp. 1370-1375, 2005.

[35] D. L. Collins *et al.*, "Design and construction of a realistic digital brain phantom," *IEEE Trans Med. Imaging*, vol. 17, no. 3, pp. 463–468, 1998.

[36] M. Watanabe *et al.*, "Performance evaluation of a high-resolution brain PET scanner using four-layer MPPC DOI detectors," *Phys. Med. Biol.*, vol. 62, no. 17, pp. 7148-7166, 2017.

[37] K. Nakaizumi *et al.*, "In vivo depiction of α7 nicotinic receptor loss for cognitive decline in Alzheimer's disease," *J. Alzheimers Dis.*, vol. 61, no. 4, pp. 1355-1365.

[38] T. Inubushi *et al*., "Neural correlates of head restraint: Unsolicited neuronal activation and dopamine release," *Neuroimage*, vol 224, pp. 117434, 2021.

[39] T. Yamaya *et al*., "Transaxial system models for jPET-D4 image reconstruction," *Phys. Med. Biol.*, vol. 50, no. 22, pp. 5339-5355, 2005.

[40] D. J. Kadrmas, "LOR-OSEM: statistical PET reconstruction from raw line-of-response histograms," *Phys. Med. Biol.*, vol. 49, no. 20, pp. 4731-4744, 2004.

[41] D. L. Snyder *et al*., "Noise and edge artifacts in maximum-likelihood reconstructions for emission tomography," *IEEE Trans. Med. Imaging*, vol. 6, no. 3, pp. 228-238, 1987.

[42] B. A. Spencer et al., "Peformance evaluation of the uEXPLORER total-body PET/CT scanner based on NEMA NU 2-2018 woth additional tests to characterize PET scanners with a long axial field of view," *J. Nucl. Med.*, vol. 62, no. 6, pp. 861-870, 2021.

[43] Y. Onishi *et al.*, "Performance evaluation of dedicated brain PET scanner with motion correction system," *Ann. Nucl. Med.*, vol. 36, no. 8, pp. 746-755, 2022.

[44] A. J. Reader and J. Verhaeghe, "4D image reconstruction for emission tomography," *Phys. Med. Biol.*, vol. 59, no. 22, R371-R418, 2014.

[45] F. Hashimoto *et al.*, "4D deep image prior: dynamic PET image denoising using an unsupervised four-dimensional branch convolutional neural network," *Phys. Med. Biol.*, vol. 66, no. 1, pp. 015006, 2021.

[46] G. Wang *et al.*, "PET parametric imaging: past, present, and future," *IEEE Trans. Radiat. Plasma Med. Sci.*, vol. 4, no. 6, pp. 663-675, 2020.

[47] A. J. Reader *et al.*, "Fully 4D image reconstruction by estimation of an input function and spectral coefficnets," *IEEE Nucl. Sci. Symp. Conf. Rec.*, Honolulu, HI, USA, pp. 3260-3266, 2007

[48] A. Rahmim *et al.*, "Motion compensation in histogram-mode and list-mode EM reconstructions: beyond the event-driven approach," *IEEE Trans. Nucl. Sci.*, vol. 51, no. 5, pp. 2588-2596, 2004.

[49] R. E. Carson *et al.*, "Design of a motion-compensation OSEM list-mode algorithm for resolution-recovery reconstruction for the HRRT," *2003 IEEE Nucl. Sci. Symp. Conf. Rec.*, Portland, OR, USA, pp. 3281-3285, 2004.

[50] M. Defrise and G. T. Gullberg, "Image reconstruction," *Phys. Med. Biol.*, vol. 51, no. 13, R139-154, 2006.

[51] Y. Onishi *et al.*, "Anatomical-guided attention enhances unsupervised PET image denoising performance," *Med. Image Anal.*, vol. 74, 102226, 2021.

[52] A. Rahmim *et al.*, "Statistical dynamic image reconstruction in state-of-the-art high resolution PET," *Phys. Med. Biol.* vol. 50, no. 20, pp. 4887-4912, 2005.

[53] A. J. Reader *et al.*, "One-pass list-mode EM algorithm for high-resolution 3-D PET image reconstruction into large arrays," *IEEE Trans. Nucl. Sci.*, vol. 49, no. 3, pp. 693-699, 2002.

[54] R. D. Badawi and P. K. Marsden, "Developments in component-based normalization for 3D PET," *Phys. Med. Biol.*, vol. 44, no. 2, pp. 571-594, 1999.

[55] C. C. Watson, "New, faster, image-based scatter correction for 3D PET," *IEEE Trans. Nucl. Sci.*, vol. 47, no. 4, pp. 1587-1594, 2000.